\documentclass[11pt,a4paper]{article}

\pdfoutput=1
\usepackage{jcappub}

\usepackage[latin1]{inputenc}
\usepackage{latexsym}
\usepackage{ifpdf}
\usepackage{pgf}
\usepackage{graphicx}
\usepackage{xy}
\usepackage{epstopdf}
\usepackage{verbatim}

\begin{document}

\title{Constraining the cosmic deceleration-acceleration transition with type Ia supernova, BAO/CMB and H(z) data}

\author[a,b]{M. Vargas dos Santos} 
\author[a]{R. R. R. Reis}
\author[a]{I. Waga}

\affiliation[a]{Instituto de F\'\i sica, Universidade Federal do Rio de Janeiro, C. P. 68528, CEP 21941-972 Rio de Janeiro, RJ,Brazil}
\affiliation[b]{Institute for Theoretical Astrophysics, University of Oslo, P.O. Box 1029 Blindern, N-0315 Oslo, Norway}

\emailAdd{vargas@if.ufrj.br}
\emailAdd{ribamar@if.ufrj.br}
\emailAdd{ioav@if.ufrj.br}

\abstract{
We revisit the kink-like parametrization of the deceleration parameter $q(z)$  \cite{ishida08}, which considers a transition, at  redshift $z_t$, from cosmic deceleration to acceleration. In this parametrization the initial, at $z \gg z_t$, value of the q-parameter is $q_i$, its final, $z=-1$, value is $q_f$ and the duration of the transition is parametrized 
by $\tau$.   By assuming a flat space geometry we obtain constraints on the free parameters of the model using recent data from type Ia supernovae (SN Ia), baryon acoustic oscillations
(BAO), cosmic microwave background (CMB) and the Hubble parameter H(z). The use of H(z) data introduces an
explicit dependence of the combined likelihood on the present value of the Hubble parameter $H_0$, allowing us to
explore the influence of different priors when marginalizing over this parameter. We also study the importance of the CMB information in the results by considering data from WMAP7, WMAP9 (Wilkinson Microwave Anisotropy Probe - 7 and 9 years) and Planck 2015.  We show that the contours and best fit do not depend much on the different CMB data used and that the considered new BAO data is responsible for most of the improvement in the results. Assuming a flat space geometry, $q_i=1/2$ and expressing the present value of the deceleration parameter $q_0$ as a function of the other three free parameters, we obtain $z_t=0.67^{+0.10}_{-0.08}$, $\tau=0.26^{+0.14}_{-0.10}$ and $q_0=-0.48^{+0.11}_{-0.13}$, at 68\% of confidence level, with an uniform prior over $H_0$. If in addition we fix $q_f=-1$, as in flat $\Lambda$CDM, DGP and Chaplygin quartessence that are special models described by our parametrization, we get $z_t=0.66^{+0.03}_{-0.04}$, $\tau=0.33^{+0.04}_{-0.04}$ and $q_0=-0.54^{+0.05}_{-0.07}$, in excellent agreement with flat $\Lambda$CDM for which $\tau=1/3$. We also obtain for flat $w$CDM, another dark energy model described by our parametrization, the constraint on the equation of state parameter $-1.22 < w < -0.78$ at more than $99\%$ confidence level.

 }


\arxivnumber{1505.03814}

\maketitle


\section{Introduction}

Since the end of last century, when type Ia supernovae (SN Ia) observations \cite{sn_accel} 
have suggested that the expansion of the universe is speeding up, we have seen a growing number of evidences confirming this cosmic acceleration. 
In fact, much progress has been done since then, but understanding the fundamental physics 
behind cosmic acceleration is still an open question and one of the main challenges of modern physics. 

Several experiments are underway and being planned and many theoretical approaches have been suggested to
investigate cosmic acceleration.
In this work we follow the so called ``kinematical approach'' \cite{kinematic,ishida08,giostri12} which is based on a parametrization of the 
deceleration parameter $q$ as a function of the redshift $z$. The advantage of using a $q(z)$-parametrization instead of, for instance, a parametrization for the equation of state, is due to its generality. In this case, we are not restricted to general relativity and the assumptions regarding dark matter and energy are minimum. We only have to assume a metric theory of gravity. Naturally, we pursue a parametrization which can describe several 
cosmological models in a wide redshift range. 

Here, our main interest is in the transition from cosmic deceleration to acceleration. We are particularly interested in the 
redshift of this transition, its duration (in redshift)  and the present value of the deceleration 
parameter. To achieve our goal we chose the kink-like $q(z)$ expression proposed in 
\cite{ishida08}, which is quite general. For the sake of simplicity we assume a flat FLRW metric. To constrain the parameters of the model we use the following kinematical tests: SN Ia, BAO/CMB and also observational Hubble data (OHD). The BAO/CMB test is based on the ratio between the comoving angular-diameter distance to the photon-decoupling redshift $d_A(z_*)$ and the ``dilaton scale'' $D_V(z_{BAO})$. It is equal to $z_{BAO}$ times the ratio $\mathcal{R}(z_*)/\mathcal{A}(z_{BAO})$, where $\mathcal{R}$ is the CMB shift parameter and $\mathcal{A}$ is the BAO scale \cite{eisenstein05}.  The OHD test was not used before to constrain our kink-like $q(z)$-parametrization. With the inclusion 
of $H(z)$ data we also investigate the influence, on the constraints of the parameters, of different priors (two gaussian and one uniform) on the present value of the Hubble parameter $H_0$.

This paper is organized as follows: in Section \ref{sec:kink}, we present the kink-like $q$-parametrization and 
discuss some of its properties. In Section \ref{sec:obs}, the cosmological tests (SN Ia, BAO/CMB and H(z)) 
we use to constrain the model parameters are presented. Finally in Section \ref{sec:results} our main results are exhibited and discussed.


\section{The model}
\label{sec:kink}

We investigate in this work flat cosmological models which deceleration parameter ($q = -\frac{a\ddot{a}}{\dot{a}^{2}}=\frac{d}{dt}\left( 
\frac{1}{H}\right) -1$), after radiation domination, has the following redshift dependence \cite{ishida08,giostri12}
\begin{equation}
q(z) = q_{f}+\frac{(q_{i}-q_{f})}{1-\displaystyle\frac{q_{i}}{q_{f}}\left( \frac{
1+z_{t}}{1+z}\right) ^{1/\tau }}.  \label{novoq}
\end{equation}
In the above expression $z_{t}$ is the transition redshift ($q(z_{t})=0$)
from cosmic deceleration ($q>0$) to acceleration ($q<0$),  $q_{i}>0$  and $
q_{f}<0$ are, respectively, the initial $(z\gg z_{t})$ and final $
(z\rightarrow -1)$ asymptotic values of the deceleration parameter. The parameter $\tau$ is
associated with the width of the transition and is related to the jerk ($j:= \frac{a^{2}\dddot{a}}{\dot{a}^{3}}$) at the transition   
\begin{equation}
\tau ^{-1}=\left( \frac{1}{q_{i}}-\frac{1}{q_{f}}\right) \left[ \frac{dq(z)}{
d\ln (1+z)}\right] _{z=z_{t}} = \left( \frac{1}{q_{i}}-\frac{1}{q_{f}}\right) j(z_t).  \label{tau}
\end{equation}
As usual, $a=1/(1+z)$ is the scale factor of the
Friedman-Robertson-Walker metric, $H_0=100h$ km/(s Mpc) is the present value of the Hubble parameter ($H = \frac{\dot{a}}{a}$) and a dot over a quantity denotes differentiation with respect to the cosmic time.

From the definitions of $q$ and $H$ we can write
\begin{equation}
H=H_{0}\exp \int_{0}^{z}\frac{ 1+q(x)}{1+x}\;  dx,  \label{hubblepar}
\end{equation}
that can be integrated by using (\ref{novoq}) to give
\begin{equation}
\left[ \frac{H(z)}{H_{0}}\right] ^{2}=\left( 1+z\right)
^{2(1+q_{i})} 
\left[ \frac{q_{i}\left( \displaystyle\frac{1+z_{t}}{1+z}\right) ^{1/\tau }-q_{f}
}{q_{i}\left( 1+z_{t}\right) ^{1/\tau }-q_{f}}\right] ^{2\tau (q_{i}-q_{f})}.
\label{hq0}
\end{equation}

We can relate the final value of the deceleration parameter, $q_f:= q(z=-1)$, to its current value $q_0:= q(z=0)$ 
\begin{equation}
q_f=\frac{q_i \left(1+z_t \right)^{1/\tau}}{1-\displaystyle\frac{q_i}{q_0}\left[1-(1+z_t)^{1/\tau}\right]}, \label{qfq0}
\end{equation}
and by substituting the above equation in (\ref{novoq}) we get
\begin{equation}
q(z)=\frac{ q_0 q_i
   \left[\left(\frac{1+z_t}{1+z}\right)^{\frac{1}{\tau
   }}-1\right]}{(1+z)^{-1/\tau } \left[q_0+(1+z_t)^{\frac{1}{\tau }}-q_i\right]-q_0}.
   \label{newq}
\end{equation}

As mentioned above, with our kink-like $q(z)$ parametrization we aim to describe the transition from a cosmic decelerated phase to an accelerated one. Therefore, since $q_i>0$ and $q_f<0$, by using Eq. (\ref{qfq0}) it is straightforward to show that the parameter $\tau$ is constrained to the interval
\begin{equation}
0<\tau<\frac{\ln(1+z_t)}{\ln \left(1-\displaystyle\frac{q_0}{q_i}\right)}.
\label{tauconstr}
\end{equation}

In most cosmological scenarios, large scale structure formation requires that at early times the universe passes through a matter dominated era in which, $H^2 \propto (1+z)^3$ and, consequently,  $q=1/2$. Since the $q$-parametrization we are working with  is designed to describe cosmic evolution starting from a matter dominated decelerated phase on we fix $q_i=1/2$ in this work. In fact, with this assumption we do not lose much generality, and are essentially excluding from our description models with a coupling in the dark sector (see  \cite{ishida08} for more on this point). We should remark that although the validity of our parametrization can be extended to high redshift ($z\lesssim1- 2\times10^3$) it is not valid during the radiation dominated era (RDE) when $q=1$. However when dealing, for instance, with BAO and CMB data, we are implicitly assuming, besides the existence of baryons, a regular RDE and that all standard assumptions for the very early universe are valid. Finally, we should also mention that several flat cosmological models investigated in the literature are special cases of the kink-like $q(z)$ parametrization given in Eq. (\ref{novoq}), as for instance: $w$CDM, $\Lambda$CDM, the DGP brane-world model \cite{dvali00}, the quartessence  Chaplygin model  \cite{kamenshchik01,bilic01,bento02,makler02} and the Modified Polytropic Cardassian model \cite{gondolo03,wang03} (see references \cite{ishida08} and \cite{giostri12} for more on this point).


\section{Observational Data}
\label{sec:obs}

In this work we derive the constraints set by three different experiments upon the parameters $z_t$, 
$\tau$ and  $q_0$. We use the following datasets: a) observations of type Ia supernovae from the 
Joint Lightcurve Analysis (JLA) of the Sloan Digital Sky Survey (SDSS-II) and the Supernova Legacy Survey (SNLS) samples \cite{betoule14}; b) observations of baryon acoustic oscillations in the SDSS DR7 \cite{padmanabhan12} and SDSS DR9 \cite{anderson13}, in the WiggleZ Survey \cite{blake11} and in the 6dF Galaxy Survey (6dFGRS) \cite{beutler11}; c) measurements of cosmic microwave background temperature anisotropy from 2015 Planck \cite{ade15} and WMAP9-year \cite{bennett13}; d) observations of the Hubble rate $H(z)$ from \cite{simon05,gaztanaga09,stern10,moresco12,delubac15,moresco15}. In the following subsections,  the statistical analysis used to deal with those observables is briefly described.


\subsection{Type Ia supernovae}
\label{sec:obs:super}

The kink-like $q$-parametrization was used in \cite{ishida08} to investigate discrepancies, reported in the literature, between 
constraints from different SN Ia datasets in a more general context. 
In \cite{giostri12} we used only one SN Ia dataset, the 288 SN Ia compilation from Kessler 
\textit{et al.} \cite{kessler09} (sample ``e'' in their paper), analyzed with two light-curve 
fitters, MLCS2k2  \cite{jha07} and SALT2  \cite{guy07}, and showed that the tension appeared to be between the fitters instead of the datasets. It has been argued in the literature (see \cite{guy10} and
references therein) that the origins of the discrepancy between the fitters are the rest frame
U band calibration and the priors in SN Ia colors, and the community seems to agree that SALT2
is more reliable in these aspects.
In this work we use the 740 SN Ia compilation from the JLA \cite{betoule14}, 
available solely for SALT2, and follow the analysis proposed in Appendix E of  \cite{betoule14}, which we are going to briefly describe in this section.

A key quantity in SN Ia investigation is the distance modulus
\begin{equation}
\mu_{th}(z;h,{\theta}) = 5 \log\left[ D_L(z;{\theta})\right] +\mu_0(h) . \label{eq:mu}
\end{equation}
Here $h$ is the Hubble constant in units of $100$ km/s Mpc$^{-1}$, ${\theta}$ denotes the set of cosmological parameters of interest other than $h$, 
\begin{equation}
\mu_0(h) = 5\log\left(\frac{10^3 c/(\mbox{km/s})}{h}\right) = 42.38-5\log h.
\end{equation}
and
\begin{equation}
D_L(z;h,{\theta})=(1+z)\int_0^z {\frac{dz'}{H(z';h,{\theta})/H_0}}\;,
\end{equation}
is the dimensionless luminosity distance (in units of the Hubble distance today), assuming
flat geometry.

In the context of SALT2, the $j$-th SN Ia distance modulus is modeled by
\begin{equation}
\mu_j = {m_B^*}_j - M_B + {\alpha} x_{1,j} -{\beta}   c_j \;,\label{eq:muSALT2}
\end{equation}
where the parameter $c$ is a measurement of the SN Ia color, $x_1$ is related to the stretch of the light curve and 
$m_B^*$ is the peak rest-frame magnitude in the $B$ band.
Its statistical uncertainty is given by
\begin{equation}
(\sigma^\mu_j)^2=(\sigma^{m^*_B}_j)^2+{\alpha^2}(\sigma^{x_1}_j)^2 +{\beta^2} 
(\sigma^c_j)^2+2{\alpha }(\sigma^{m^*_B x_1}_j)-2{ \beta}(\sigma^{m^*_Bc}_j)-2{\alpha  \beta}(\sigma^{x_1c}_j)\;.\label{eq:sigmaSALT2}
\end{equation}
Here the quantities $(\sigma^{m^*_B}_j)^2$, $(\sigma^{x_1}_j)^2$, $(\sigma^c_j)^2$, $\sigma^{m^*_B x_1}_j$, $\sigma^{m^*_Bc}_j$ and $\sigma^{x_1c}_j$ are the components of the covariance matrix of $(m^*_B, x_1, c)$. We remark that $m_B^*$, $x_1$ and $c$ are derived from the fit to the light curves while the global parameters $\alpha$, $\beta$ and the absolute magnitude $M$ are estimated simultaneously with the cosmological parameters and are marginalized over when obtaining the cosmological parameter constraints.

In order to take into account the dependence of the absolute magnitude with host galaxy properties, it was introduced an additional parameter $\Delta M$ (originally proposed by \cite{conley11}) such that
\begin{eqnarray}
M_B  = & M_B^1 & \hbox{if}\quad M_{stellar} < 10^{10} M_{\odot}, \nonumber \\
M_B  = & M_B^1 + \Delta_M & \hbox{otherwise},
\end{eqnarray}
where $M_{stellar}$ is the host galaxy stellar mass.

The likelihood for SN Ia data is given by
\begin{equation}
L_{SN}=\frac{1}{\sqrt{(2\pi)^N\hbox{det}\,{C}}}\exp(-
\chi^2_{SN}/2)\,,
\label{like}
\end{equation}
where $C$ is the $N\times N$ covariance matrix of the $N$ SN Ia observations. 

In this work we assume that the estimate of distances is independent of the estimate of cosmological parameters  \cite{betoule14}. 
Under this assumption, the distance modulus can be approximately written as
\begin{equation}
\bar{\mu}(z)=(1-\alpha)\mu_b+\alpha\mu_{b+1},
\label{eq:binneddist}
\end{equation}
where $\alpha=\log(z/z_b)/\log(z_{b+1}/z_b)$ and $\mu_b$ is the distance modulus at $z_b$.

This function was fitted to the measured Hubble diagram by minimizing the $\chi^2$ function
\begin{equation}
\chi^2_{SN}=(\mu-\bar{\mu}(z))^{t}C^{-1}(\mu-\bar{\mu}(z)).
\end{equation}
The parameters to be determined are $\alpha$, $\beta$, $\Delta_M$ and the values of $\mu_b$ in each bin (they chose 31 log-spaced points $z_b$ in the redshift range $0.01<z<1.3$). In order to obtain unique values for $\mu_b$ it was used a fiducial value of $M_B^1=-19.5$.

Once the correlation matrix $C_b$ for the parameters $\mu_b$ is obtained we can use it in the cosmological analysis by inserting in Eq. (\ref{like}) the following function
\begin{equation}
\chi^2_{SN}=(\mu_b-\mathcal{M}-5\log D_L(z_b,\mathbf{\theta}))^{t}C_b^{-1}(\mu_b-\mathcal{M}-5\log D_L(z_b,\mathbf{\theta})),
\label{eq:chi2sne}
\end{equation}
where $\mathcal{M} = M - 5 \log h +42.38$. The cosmological constraints are obtained marginalizing over the parameter $\mathcal{M}$ with an uniform prior. We remark that in our analysis the full covariance matrices, including systematic errors, as reported by \cite{betoule14}, were used.


\subsection{Baryon acoustic oscillations and cosmic microwave background}
\label{sec:obs:BAO}

In this section we are going to briefly review the combined BAO/CMB test. However, before doing so, some remarks are required. We emphasise that our approach is kinematical in which we avoid strong assumptions about the dark sector. All the tests we consider here are essentially based only on distances and we are not considering tests directly based on the growth of perturbations. Furthermore, as mentioned before, although the parametrization given in Eq. (\ref{novoq}) is not assumed to be valid during RDE, we assume, besides the existence of baryons, a regular RDE and that all standard assumptions for the very early universe are valid. Thus, we can define the comoving sound horizon at the photon-decoupling epoch by
\begin{equation}
r_s(z_*)= \frac{c}{\sqrt{3}} \int_0^{1/(1+z_*)}\frac{da}{a^2H(a)\sqrt{1+(3\Omega_{b0} / 4 \Omega_{\gamma 0})a} },\label{eq:r_s}
\end{equation}
where $\Omega_{\gamma 0}$ and $\Omega_{b0}$ are, respectively, the present value of the photon and baryon density parameter and $z_*$ is the redshift of photon decoupling. To obtain the BAO/CMB constraints another relevant quantity is the redshift of the drag epoch ($z_d$), when the photon pressure is no longer able to avoid gravitational instability of the baryons. The WMAP9 \cite{bennett13} values for these two redshifts are $z_*=1090.97^{+0.85}_{-0.86}$ and $z_d=1020.7\pm 1.1$, while the 2015 results for Planck \cite{ade15} are $z_*=1090.00\pm 0.29$ and $z_d=1059.62\pm 0.31$. 
Also important are the ``acoustic scale''
$l_A=\pi\frac{d_A(z_*)}{r_s(z_*)}$, and  the ``dilation scale'',  
$D_V(z):=\left[ d_A^2(z) cz/H(z) \right]^{1/3}$, introduced in \cite{eisenstein05}. Here,  $d_A(z_*)= c \int_{0}^{z_*}dz'/H(z')$ is the comoving angular-diameter distance. 

We are using the results from Padmanabhan et al. \cite{padmanabhan12}, which reported a measurement of $r_s/D_V$ at $z=0.35$, ($r_s(z_d)/D_V(0.35)=0.1126\pm0.0022$), Anderson et al. \cite{anderson13}, ($r_s(z_d)/D_V(0.57)=0.0732\pm0.0012$), Beutler et al. \cite{beutler11}, ($r_s(z_d)/D_V(0.106)=0.336\pm0.015$) and Blake et al. \cite{blake11}, which obtained results at $z=0.44$, $z=0.6$ and $z=0.73$ ($r_s(z_d)/D_V(0.44)=0.0916\pm0.0071$, $r_s(z_d)/D_V(0.6)=0.0726\pm0.0034$ and $r_s(z_d)/D_V(0.73)=0.0592\pm0.0032$).

We combine theses results with the Planck 2015 \cite{ade15} values $100\theta_*=1.04106\pm 0.00031$, which gives 
$l_A = \pi/\theta_*= 301.77\pm0.09$, $r_s(z_d)=147.41\pm 0.30$ and
$r_s(z_*)=144.71\pm 0.31$, for the combined analysis TT, TE, EE+lowP+lensing (see Table 4 of that
reference). Following \cite{ade14}, we multiply the value of $r_s(z_d)$ by a constant factor
of 1.0275 to take into account the difference between CAMB \cite{camb} results and the Eisenstein and Hu 
approximation \cite{eisenstein98} used in the BAO analysis.
With this information we can write the $\chi^2$ for the BAO/CMB as
\begin{equation}
\chi^2_{BAO/CMB}={\bf X^tC^{-1}X},
\label{chi2baocmb}
\end{equation}
where
\begin{equation}
{\bf X}=\left(
          \begin{array}{cccc}
          \displaystyle\frac{d_A(z_*)}{D_V(0.106)} -30.84 \\
            \displaystyle\frac{d_A(z_*)}{D_V(0.35)} -10.33 \\
            \displaystyle\frac{d_A(z_*)}{D_V(0.57)} -6.72 \\
             \displaystyle\frac{d_A(z_*)}{D_V(0.44)} -8.41 \\
             \displaystyle\frac{d_A(z_*)}{D_V(0.6)} -6.66 \\
              \displaystyle\frac{d_A(z_*)}{D_V(0.73)} -5.43 \\
          \end{array}
        \right)
\end{equation}
and
\begin{equation}
{\bf C^{-1}}=\left(
          \begin{array}{cccccc}
0.52552	& -0.03548	& -0.07733	& -0.00167	& -0.00532 &	-0.00590 \\
-0.03548	& 24.97066	& -1.25461	& -0.02704	& -0.08633 &	-0.09579 \\
-0.07733	& -1.25461	& 82.92948	& -0.05895	& -0.18819 &	-0.20881 \\
-0.00167	& -0.02704	& -0.05895	& 2.91150	& -2.98873 &	1.43206 \\
-0.00532	& -0.08633	& -0.18819	& -2.98873	& 15.96834 &	-7.70636 \\
-0.00590	& -0.09579	& -0.20881	& 1.43206	& -7.70636 &	15.28135  \\          
\end{array}
        \right)
\end{equation}
is the inverse covariance matrix derived by using the equations given above together with the correlation coefficients $r=0.39$, $r=0.453$ and $r=-2.12417\times 10^{-7}$ calculated for the  $r_s/D_V$  pair of measurements at $z=(0.44, 0.6)$, $z=(0.6, 0.73)$ and $z=(0.44, 0.73)$ as presented in \cite{hinshaw13}. In Ref. \cite{giostri12} the correlation between the measurements at $z=0.44$ and
$0.73$ was assumed to be zero since it was not available in the original reference \cite{blake11}.

We also used the results from WMAP9-year \cite{bennett13}: 
$l_A = 302.35\pm0.65$, $r_s(z_d)=152.3\pm 1.3$ and $r_s(z_*)=145.8\pm 1.2$ 
which gives
\begin{equation}
{\bf X}=\left(
          \begin{array}{cccc}
          \displaystyle\frac{d_A(z_*)}{D_V(0.106)} -30.96 \\
            \displaystyle\frac{d_A(z_*)}{D_V(0.35)} -10.37 \\
            \displaystyle\frac{d_A(z_*)}{D_V(0.57)} -6.74 \\
             \displaystyle\frac{d_A(z_*)}{D_V(0.44)} -8.44 \\
             \displaystyle\frac{d_A(z_*)}{D_V(0.6)} -6.69 \\
              \displaystyle\frac{d_A(z_*)}{D_V(0.73)} -5.45 \\
          \end{array}
        \right)
\end{equation}
and
\begin{equation}
{\bf C^{-1}}=\left(
          \begin{array}{cccccc}
0.50557	& -0.29215	& -0.63687	& -0.01373	& -0.04382 &	-0.04862 \\
-0.29215	& 20.60694	& -10.33210	& -0.22271	& -0.71095 &	-0.78883 \\
-0.63687	& -10.33210	& 62.47146	& -0.48551	& -1.54983 &	-1.71962 \\
-0.01373	& -0.22271	& -0.48551	& 2.87954	& -2.99476 &	1.38827 \\
-0.04382	& -0.71095	& -1.54983	& -2.99476	& 15.74975 &	-7.75021 \\
-0.04862	& -0.78883	& -1.71962	& 1.38827	& -7.75021 &	15.04645 \\
          \end{array}
        \right)
\end{equation}


\subsection{H(z)}
\label{sec:obs:H}

Constraining cosmological models with observational Hubble data (OHD), i.e., by direct measurements of the expansion rate, has received considerable attention in recent years. The two main methods to measure the Hubble parameter as a function of redshift, $H(z)$, are the \textit{differential age method}  and the \textit{radial BAO size method}. For review, including a discussion on systematic uncertainties and possible source of errors, see \cite{zhang10}. 

The differential age method has been proposed in \cite{jimenez12}. It has been shown in this reference, that by measuring the age differences between two ensembles of passively evolving galaxies at somewhat different redshifts, it is possible to obtain the derivative $dz/dt$ and this quantity is directly related to $H(z)$,
\begin{equation}
H(z)=-\frac{1}{1+z}\frac{dz}{dt}.
\end{equation}
By using the radial BAO peak position as a standard ruler in the radial direction, in Ref. \cite{gaztanaga09}, for the first time, it has been obtained a direct measurement of $H(z)$ by the second method mentioned above. 
Recently, Farooq and Ratra \cite{farooq13} compiled a list of 28 independent measurements of the Hubble parameter at different redshifts and used these measurements to constrain cosmological models. 
Here we use their compilation (see Table 1 of \cite{farooq13}), but substitute the result 
from \cite{busca12} (at $z=2.3$) with that from Delubac \textit{et al.} \cite{delubac15} (at $z=2.34$). In \cite{delubac15}  it was used 
over 137500 quasars from Data Release 11 of the SDSS-III Baryon Oscillation Spectroscopic Survey (BOSS), more than twice
the number of quasars considered in \cite{busca12}. This is the most important improvement of \cite{delubac15} over \cite{busca12}. Other treatment improvements are described in \cite{delubac15}. To avoid possible correlations and to assure data independence, we also excluded the three data points from \cite{blake12}, at $z=0.44$, $z=0.60$ and $z=0.68$, since the same data is used in our CMB/BAO analysis \cite{blake11}. In our compilation we have also included two new measurements from cosmic chronometers at redshifts $1.363$ and $1.965$ from \cite{moresco15}. 

In order to constrain the $\mathbf{\theta}$ parameters of the model we use the following $\chi_{H}^{2}$ function

\begin{equation}
        \chi_{H}^{2} = \sum_{i=1}^{27} \left[ \frac{H_{th}(z_i, H_{0}, \mathbf{\theta})-H_{obs,i}}{\sigma_{H,i}}\right]^2,
        \label{chi_hub}
\end{equation}
where $H_{th}(z_i, H_{0}, \mathbf{\theta})$ denotes the Hubble parameter at $z_i$ predicted by the models, $H_{obs,i}$ is the i-th measured one and $\sigma_{H,i}$ is its uncertainty.

In this work we marginalize $ \chi_{H}^{2}$ over $H_0$ (see the Appendix for details), using three different priors: 
\begin{enumerate}
\item an uniform prior over all $H_0$ possible values.
\item A Gaussian prior using the measurement from Riess \textit{et al.} 
\cite{riess11}, $H_0=73.8 \pm 2.4$ km s$^{-1}$ Mpc$^{-1}$;
\item A Gaussian prior using the measurement from Aubourg \textit{et al.} \cite{aubourg14}, $H_0=67.3\pm1.1$ km s$^{-1}$ Mpc$^{-1}$.
\end{enumerate}


\section{Results and discussion}
 \label{sec:results}
 
 Since the SN Ia, BAO/CMB and H(z) data are independent from each other we can write the combined $\chi^2$ 
statistics as (see the Appendix)
\begin{equation}
   \chi^2 = \chi^2_{SN,marg} + \chi^2_{BAO/CMB} +\chi^2_{H,marg}\;.
   \label{eq:chi2}
\end{equation}
where $\chi^2_{SN,marg}$, $ \chi^2_{BAO/CMB}$ and $\chi^2_{H,marg}$ are given, respectively, by Eqs.  (\ref{snmarg}),  (\ref{chi2baocmb}) and (\ref{Hmarg}). At this point it should be pointed out that on any occasion when when we refer to an uniform prior marginalization over a parameter, we mean a marginalization over all its possible values.

\begin{figure*}[tbp]
\begin{center}
\includegraphics[scale=0.3]{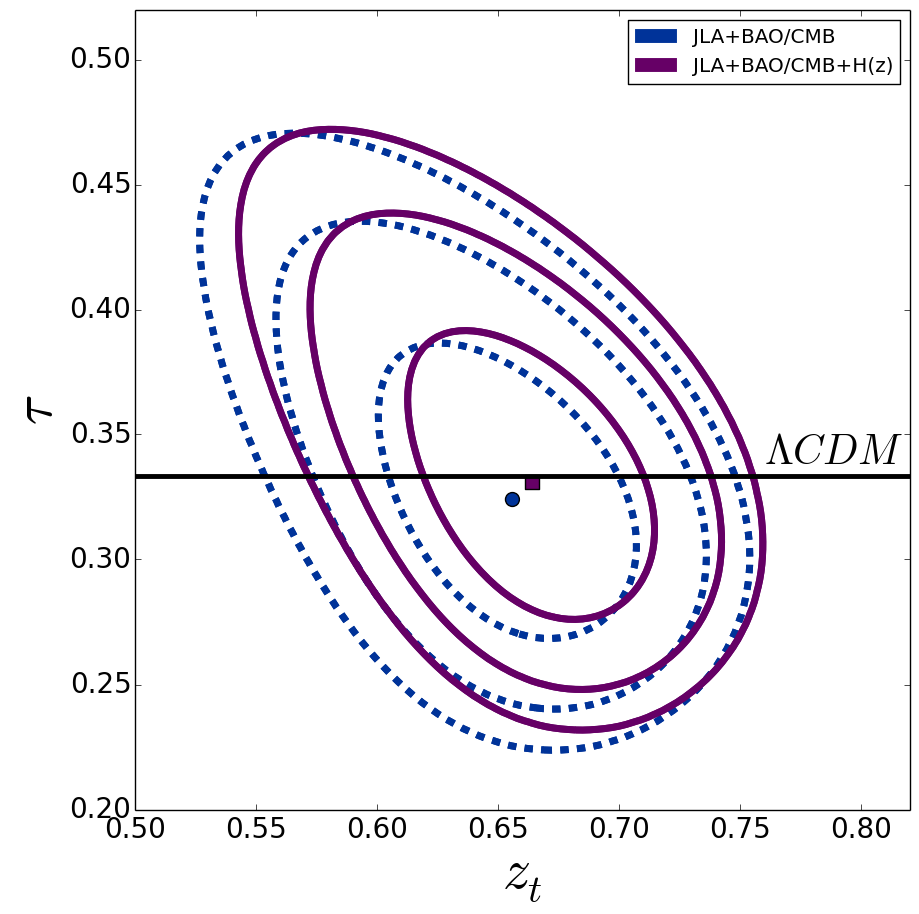}\\
\includegraphics[scale=0.3]{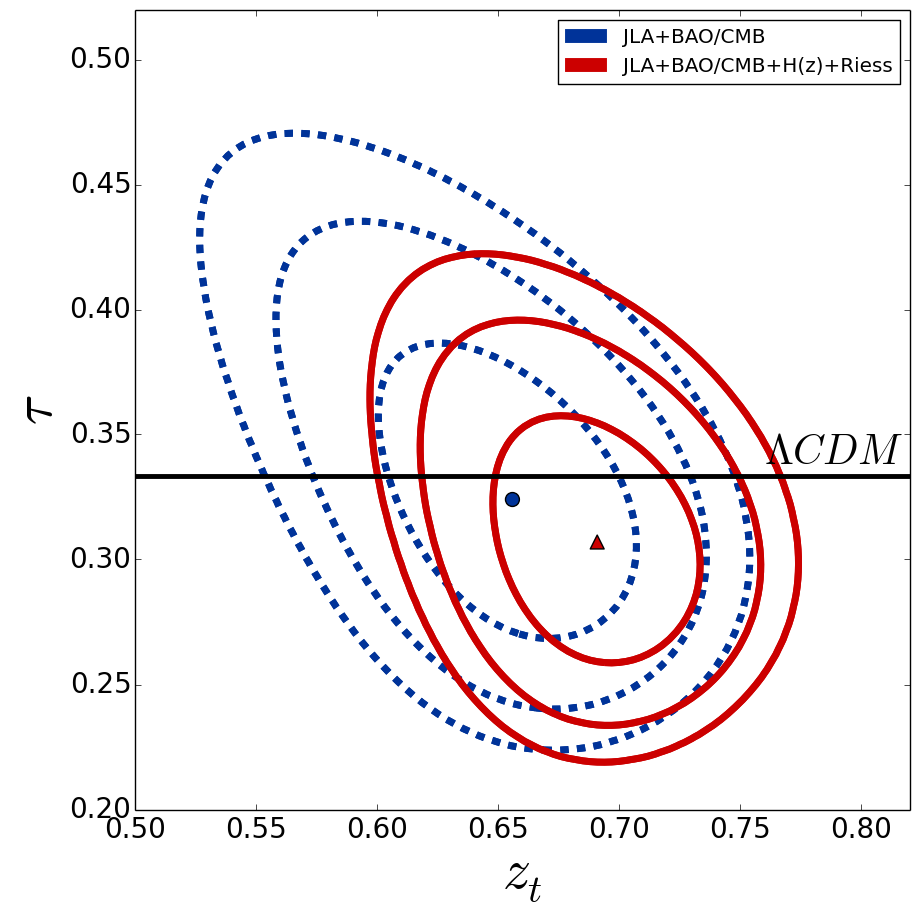}
\includegraphics[scale=0.3]{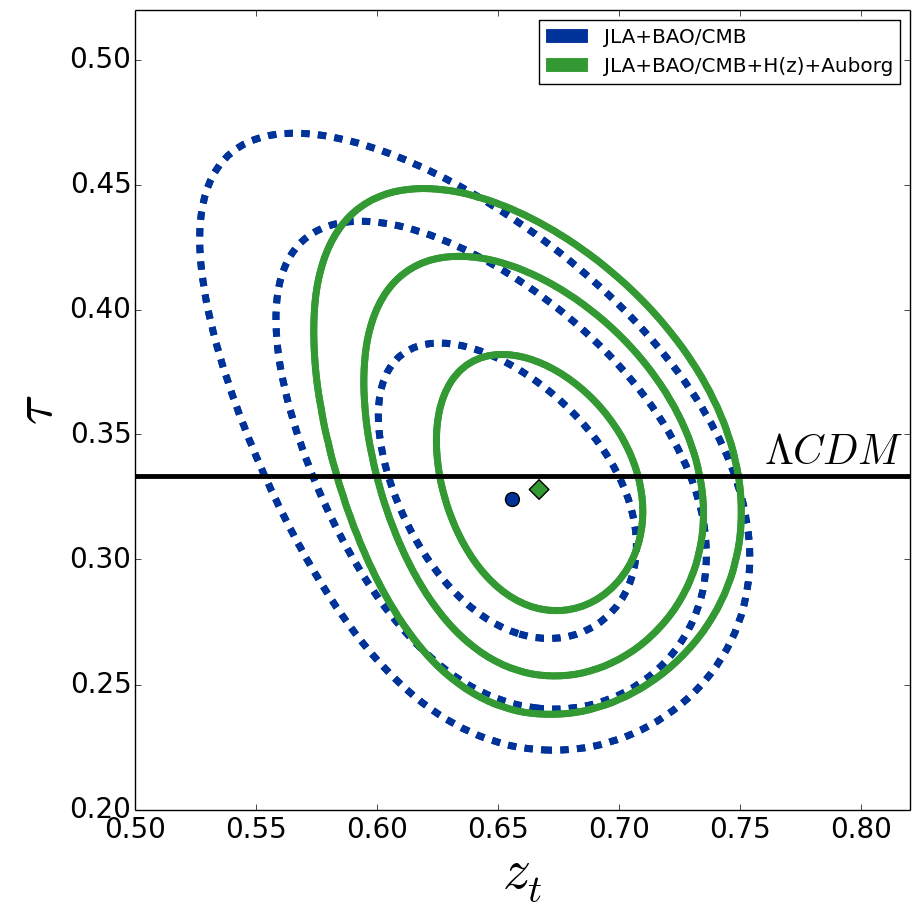}
\end{center}
\caption{\small{68\%, 95\% and 99\% confidence contours in the plane $z_t$ vs. $\tau$, for
$q_i=0.5$ and $q_f=-1$. In all panels the blue dashed lines are the contours for SN Ia + BAO/CMB(Planck). In the top panel the purple lines are the contours for SN Ia + BAO/CMB(Planck) + H(z), marginalized over $H_0$ with an uniform prior. In the bottom left panel the red lines are the contours for the same data, marginalized using the Riess \textit{et al.} measurement as a Gaussian prior. In the bottom right panel the green lines 
are for the same data using the Aubourg \textit{et al.} measurement as prior. In all panels the horizontal line corresponds to $\Lambda$CDM ($\tau=1/3$) and the small dots and squares indicate the best-fit values. }}
\label{fig1}
\end{figure*}

Let us first consider the case in which, besides fixing $q_i=1/2$, the final value of the deceleration parameter is fixed to the value $q_f = -1$, such that asymptotically in the future ($z=-1$) a de Sitter phase is reached. Popular cosmological models like flat $\Lambda$CDM, quartessence Chaplygin, DGP etc, belong to this class of models as discussed in Ref. \cite{giostri12}. 
In Fig. \ref{fig1} we show the $68\%$, $95\%$ and $99\%$ confidence level regions in the $(z_t, \tau)$ plane.
In all panels of this figure the blue dashed lines are contours for SN Ia + BAO/CMB(Planck) data. In the upper panel we also show the contours for SN Ia + BAO/CMB(Planck) + H(z) data, marginalizing over $H_0$ with an uniform prior (purple lines). We see that including the $H(z)$ data, assuming an uniform prior over $H_0$, does not change very much the contours nor the best fit. In the bottom panels we also show the results for the SN Ia + BAO/CMB(Planck) + H(z) data, but marginalized over $H_0$ with Gaussian priors. In the bottom left panel we used as prior for $H_0$, the Riess \textit{et al.} \cite{riess11} measurement (red lines) and in the bottom right we used the Aubourg \textit{et al.} \cite{aubourg14} result (green lines). Now the constraints are tighter, making clear the important role of the $H_0$ prior when including  $H(z)$ data. We obtain that using the Riess \textit{et al.} result pushes the contours for lower values of $\tau$ and higher values of $z_t$ when compared to the other two cases. We can quantify the improvement with respect to the previous work \cite{giostri12} by comparing the areas inside the $95\%$ C.L. contours. The areas ratio between the above results and the one from Giostri \textit{et al.} \cite{giostri12}, that used WMAP7, are $0.23$, $0.23$ and $0.29$ for the Riess, Aubourg and uniform priors. If we do not consider the H(z) data but use Planck instead of WMAP7 we get $0.31$ for the areas ratio. It is also worth to mention that flat $\Lambda$CDM models ($\tau = 1/3$), represented by the black horizontal line in all figures, are in excellent accordance with the data in all cases analyzed.

\begin{figure*}[tbp]
\begin{center}
\includegraphics[scale=0.3]{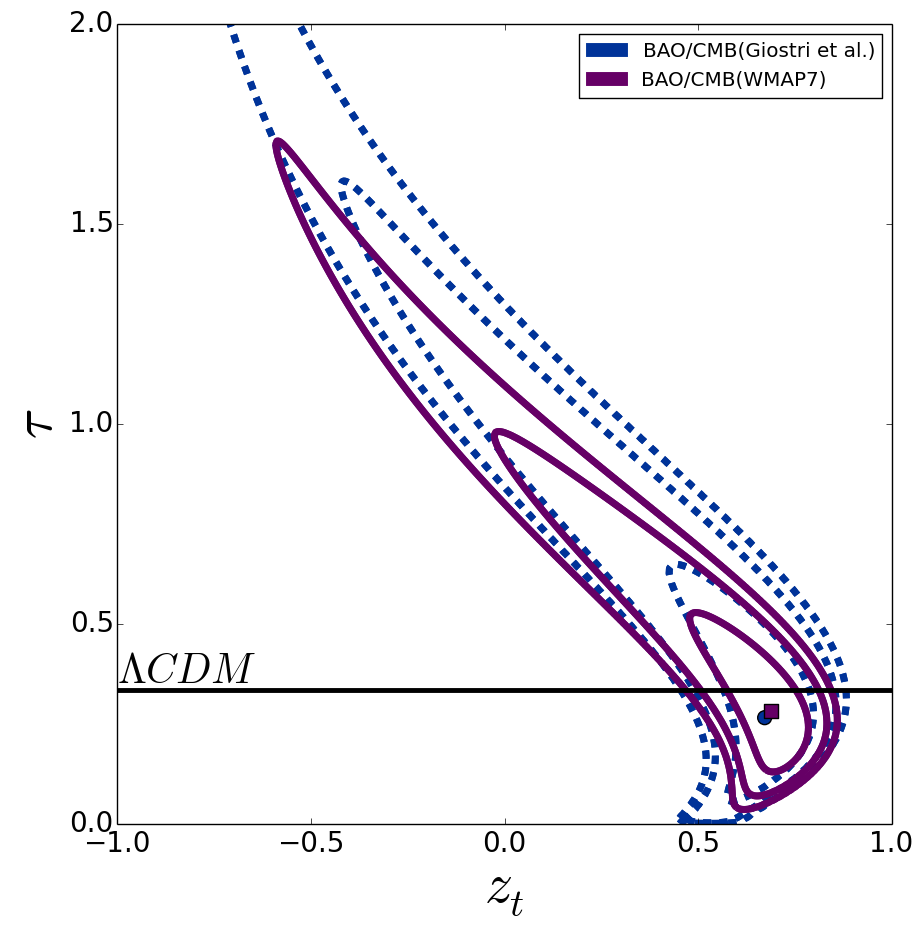}
\includegraphics[scale=0.3]{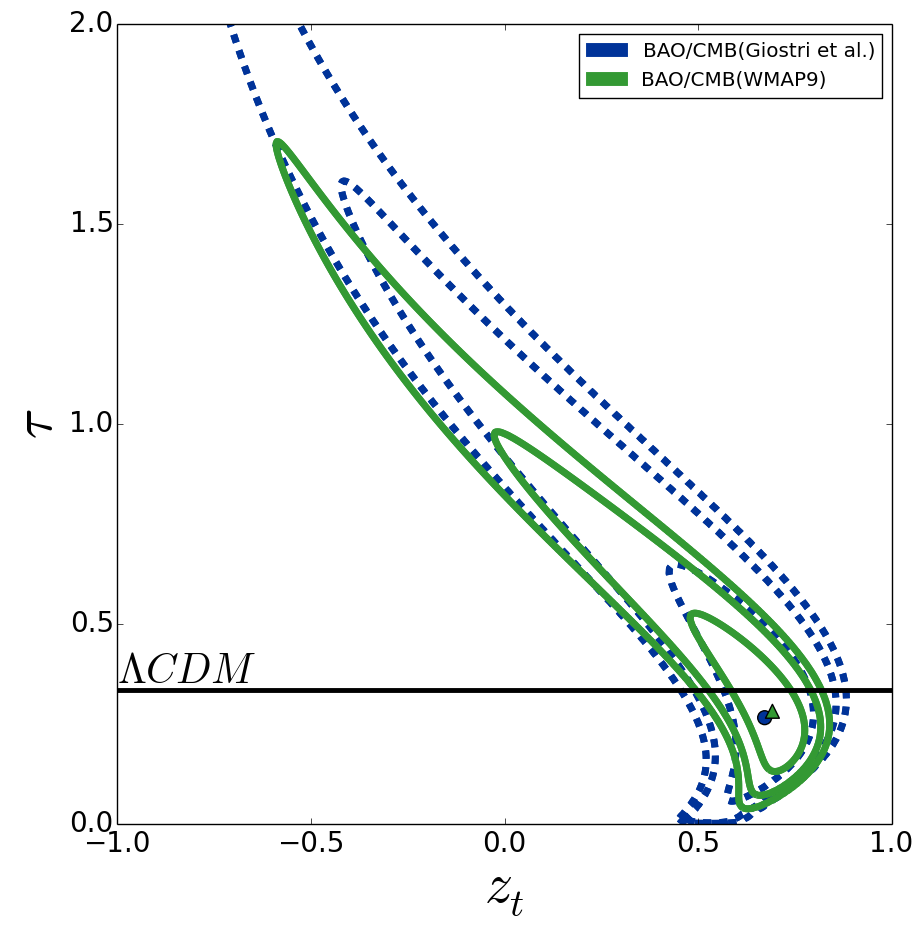}
\includegraphics[scale=0.3]{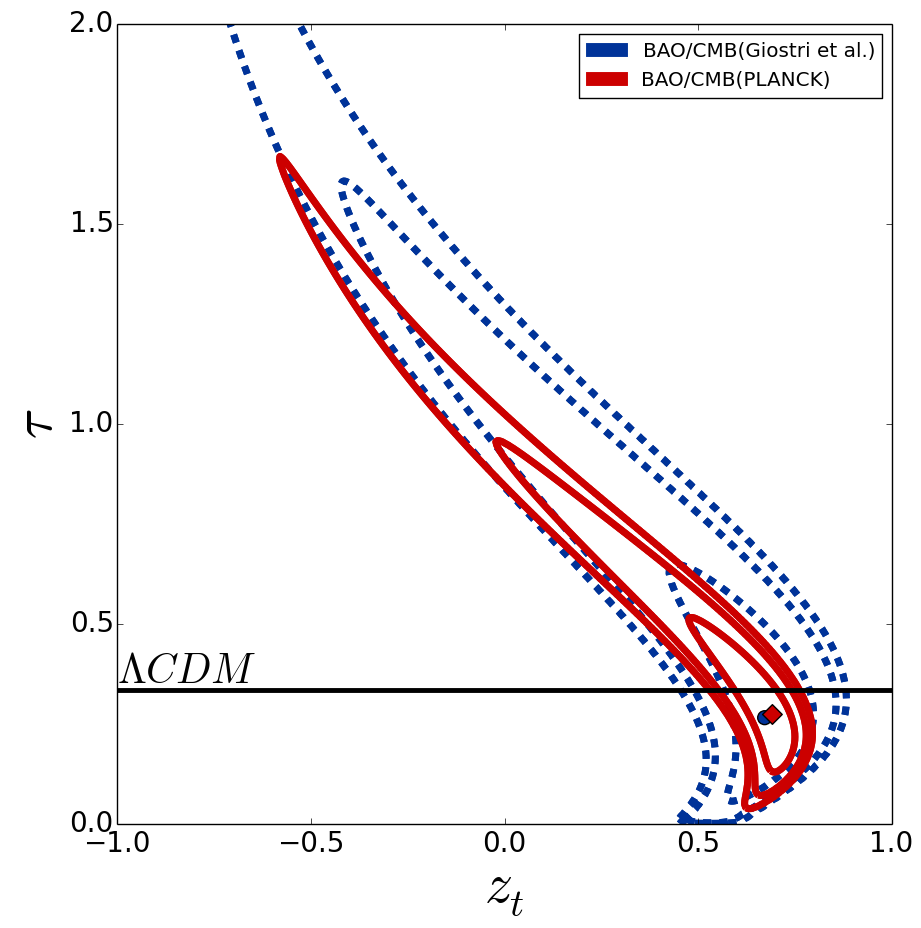}
\end{center}
\caption{\small{68\%, 95\% and 99\% confidence contours in the plane $z_t$ vs. $\tau$, for $q_i=0.5$ and $q_f=-1$, considering only BAO/CMB data. In all panels the blue dashed lines are the contours for BAO/CMB(WMAP7) data from Giostri \textit{et al.} \cite{giostri12}. We show the results for our analysis, which uses a different BAO dataset, combined with CMB information from WMAP7 (in the top left panel, the purple lines), WMAP9 (in the top right panel the green lines) and Planck (in the bottom panel, the red lines). In all panels the horizontal line corresponds to $\Lambda$CDM ($\tau=1/3$)}}
\label{fig2}
\end{figure*}

It is important to highlight the significant role of the new BAO data in the improvement of the constraints when considering the BAO/CMB test alone. The difference in the BAO dataset used in this work from that used in \cite{giostri12} (see Table 1 in this reference) is that here we dropped the measurements from \cite{percival10} at $z=0.2$ (which has one of the largest errors) and at $z=0.35$. We considered instead the more precise estimates at $z=0.35$ from \cite{padmanabhan12} and at $z=0.57$ from \cite{anderson13}. 
In Fig.\,\ref{fig2} we compare our results for this test to the
Giostri \textit{et al.}  \cite{giostri12} result, using CMB information from WMAP7 (the same used in \cite{giostri12}), WMAP9 and Planck 2015 combined with the new BAO data. In the top left panel the only difference between the Giostri \textit{et al.}
result (blue dashed lines) and ours (purple lines) is the BAO data. We can see in the top right and bottom panels the same analysis replacing the WMAP7 information by WMAP9 (green lines) and Planck 2015 (red lines). 
We can quantify the improvement calculating the areas inside the contours. The  95\% C.L. areas ratio relative to the Giostri \textit{et al.} result are 0.29 for Planck, 0.40 for WMAP9 and
0.47 for WMAP7. The numbers are similar for 68\% and 99\% C.L.. Therefore, we conclude that the new BAO data gives a major contribution to the improvement found. This indicates that more precise BAO measurements are very important to achieve tighter constraints on the parameters.

\begin{figure*}[tbp]
\begin{center}
\includegraphics[scale=0.5]{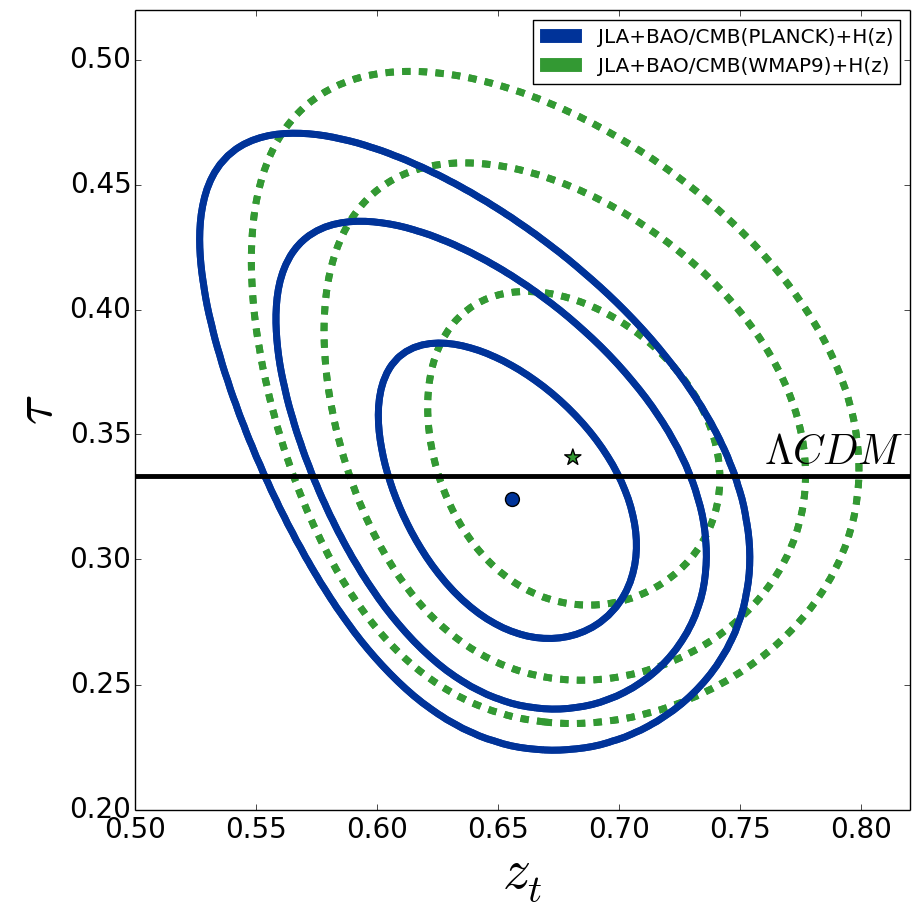}
\end{center}
\caption{\small{68\%, 95\% and 99\% confidence contours in the plane $z_t$ vs. $\tau$, for $q_i=0.5$ and $q_f=-1$, using the combined data SN Ia + BAO/CMB + H(z), marginalized over $H_0$ with an uniform prior. For the green dashed lines we used WMAP9 information while for the blue lines we used Planck 2015. The horizontal line corresponds to $\Lambda$CDM ($\tau=1/3$) and the small dot and square indicate the best-fit values.}}
\label{fig3}
\end{figure*}

We show in Fig. \ref{fig3} a comparison between the constraints for the case SN Ia + BAO/CMB + H(z), marginalized over $H_0$ with an uniform prior, using WMPA9 (green dashed lines) and Planck (blue lines). We can see that Planck's higher precision gives us more stringent constraints. The 95\% C.L. contour area ratio with Planck with respect to the one with WMAP9 is 0.71. We verified that this property also holds when the two other $H_0$ priors are used. In the following we base our analyses by considering only the Planck information for CMB.

We now consider the situation in which $q_f$ is free and can assume any value (we still fix $q_i=1/2$). In this very 
general case, the confidence regions of three model parameters ($z_t$, $\tau$ and $q_f$) have to be determined. However, following Ref. \cite{giostri12}, instead of working with $q_f$, we use $q_0$, the present value of the deceleration parameter.  
As discussed in Sec. \ref{sec:kink}, Eq. (\ref{qfq0}) gives $q_f$ in terms of $z_t$, $\tau$ and $q_0$ allowing us to re-express $q(z)$ in terms of  these parameters (see Eq. (\ref{newq})). 
It is important to mention that in the $q_f$ free case, to obtain the confidence regions, following  \cite{giostri12}, we explicitly used the restriction on $\tau$ given by Eq. (\ref{tauconstr}).  As discussed in Section \ref{sec:kink}, this restriction on $\tau$ expresses the fact that the transition in the universe is from a decelerated phase to an accelerated one.

\begin{figure*}[tbp]
\begin{center}
\includegraphics[scale=0.32]{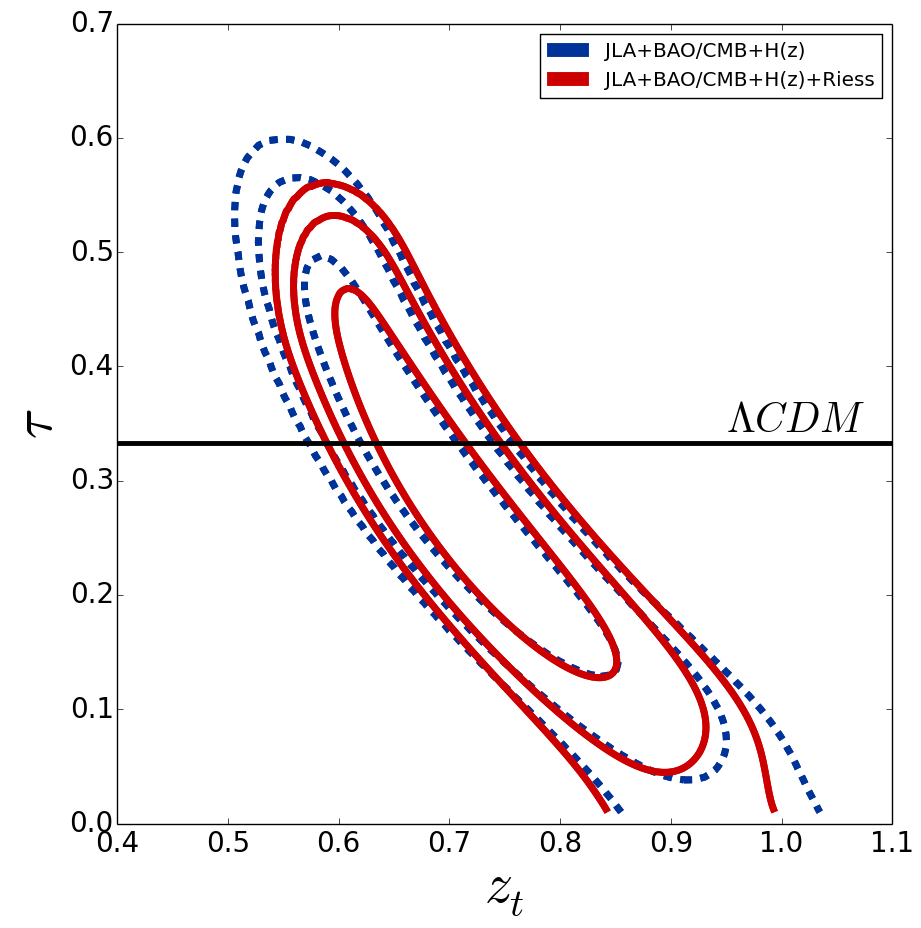}
\includegraphics[scale=0.32]{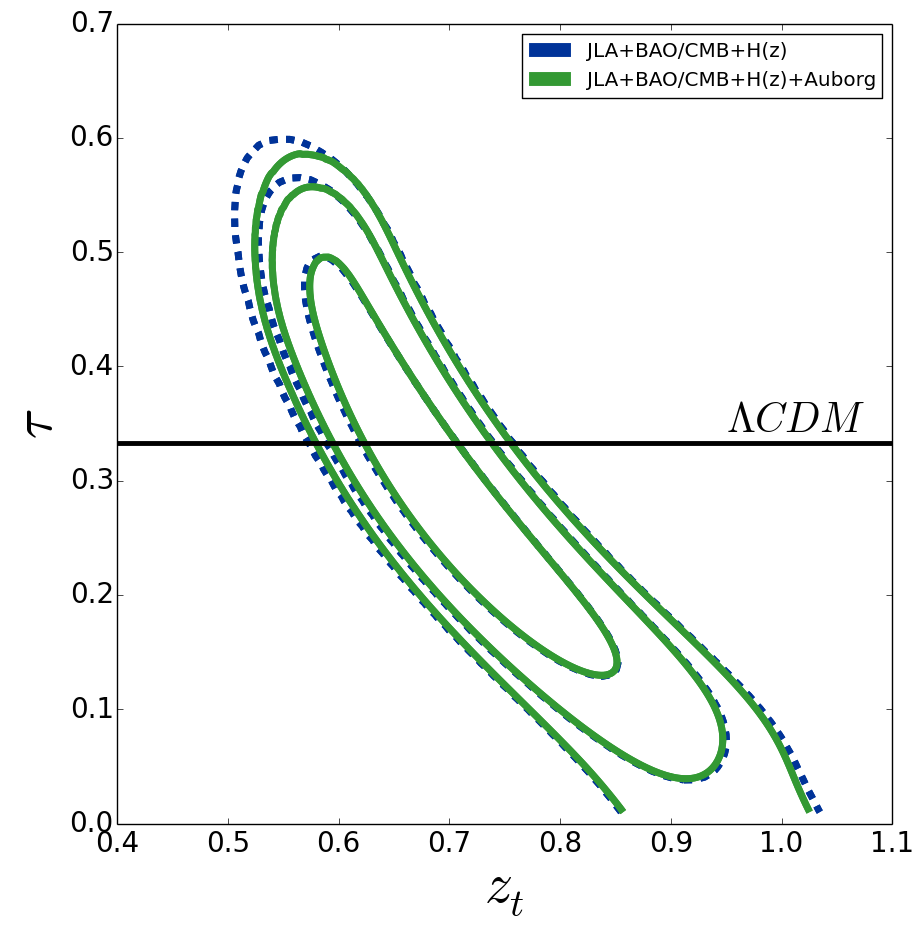}
\end{center}
\caption{\small{68\%, 95\% and 99\% confidence contours in the plane $\tau$ vs. $z_t$ for SN Ia + BAO/CMB(Planck) + H(z). The blue dashed lines in both panels are the contours for the case of marginalization over $H_0$ with an uniform prior while the red ones (left panel) are for the Riess \textit{et al.} prior and the green ones are for the Auborg \textit{et al.} prior. In all cases, the marginalization over the third model parameter is done with an uniform prior. In both panels the horizontal dashed line corresponds to $\Lambda$CDM ($\tau=1/3$).}}
\label{fig4}
\end{figure*}

\begin{figure*}[tbp]
\begin{center}
\includegraphics[scale=0.32]{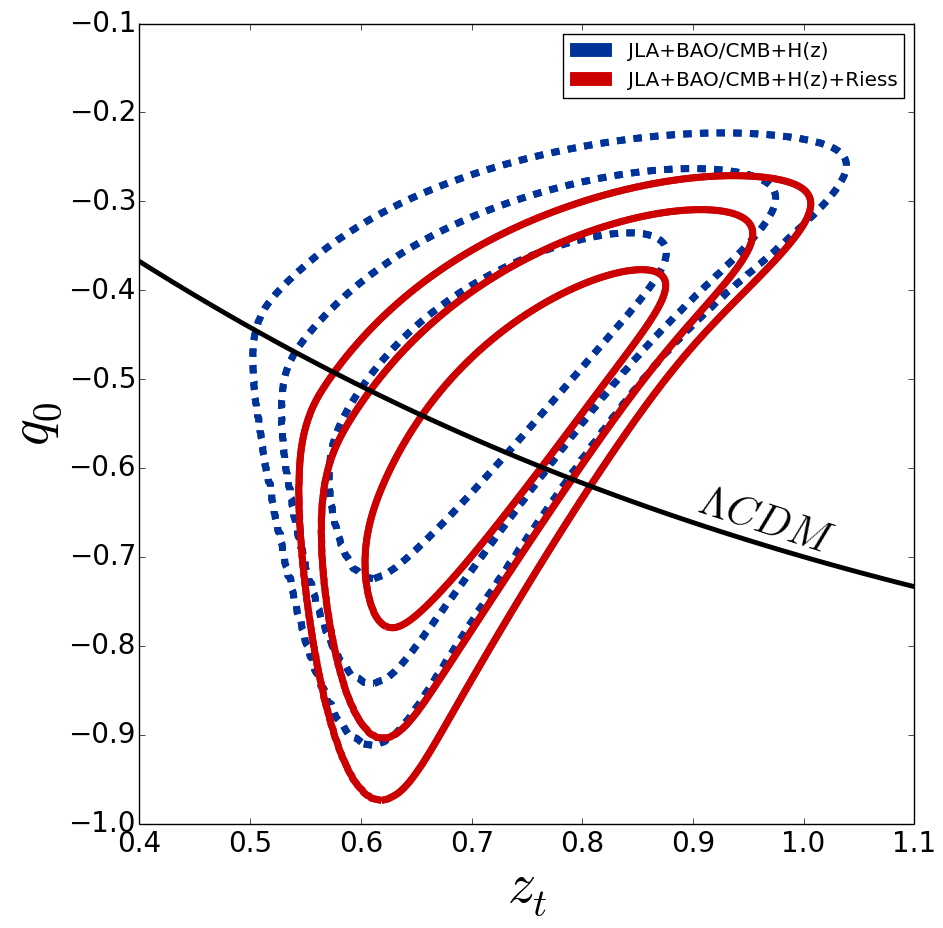}
\includegraphics[scale=0.32]{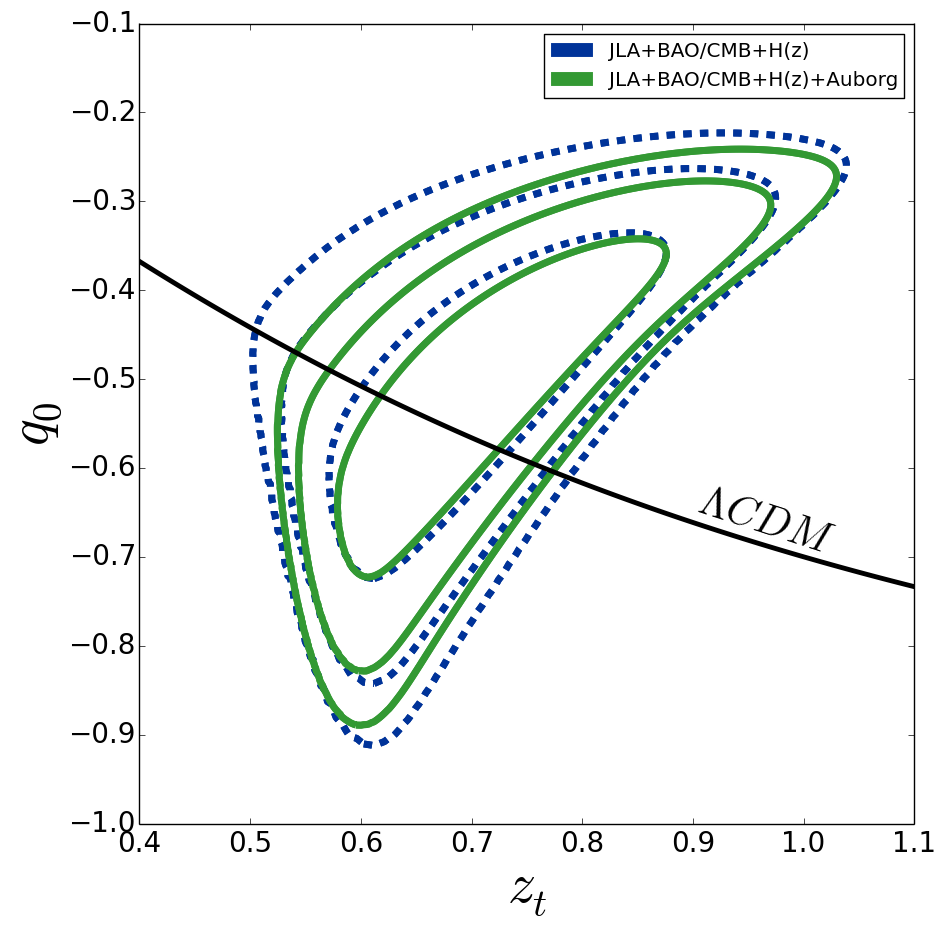}
\end{center}
\caption{\small{68\%, 95\% and 99\% confidence contours in the plane $q_0$ vs. $z_t$
for SN Ia + BAO/CMB(Planck) + H(z). The lines and color conventions are the same described in  Fig. \ref{fig4}}}
\label{fig5}
\end{figure*}

\begin{figure*}[tbp]
\begin{center}
\includegraphics[scale=0.32]{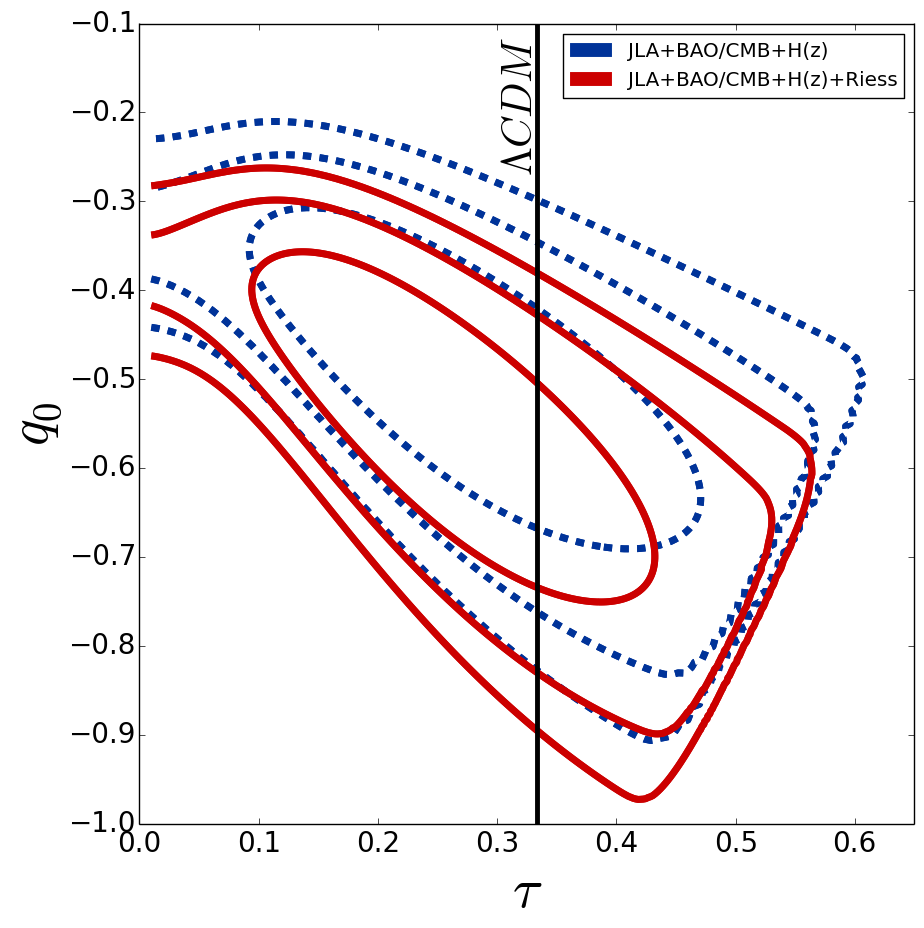}
\includegraphics[scale=0.32]{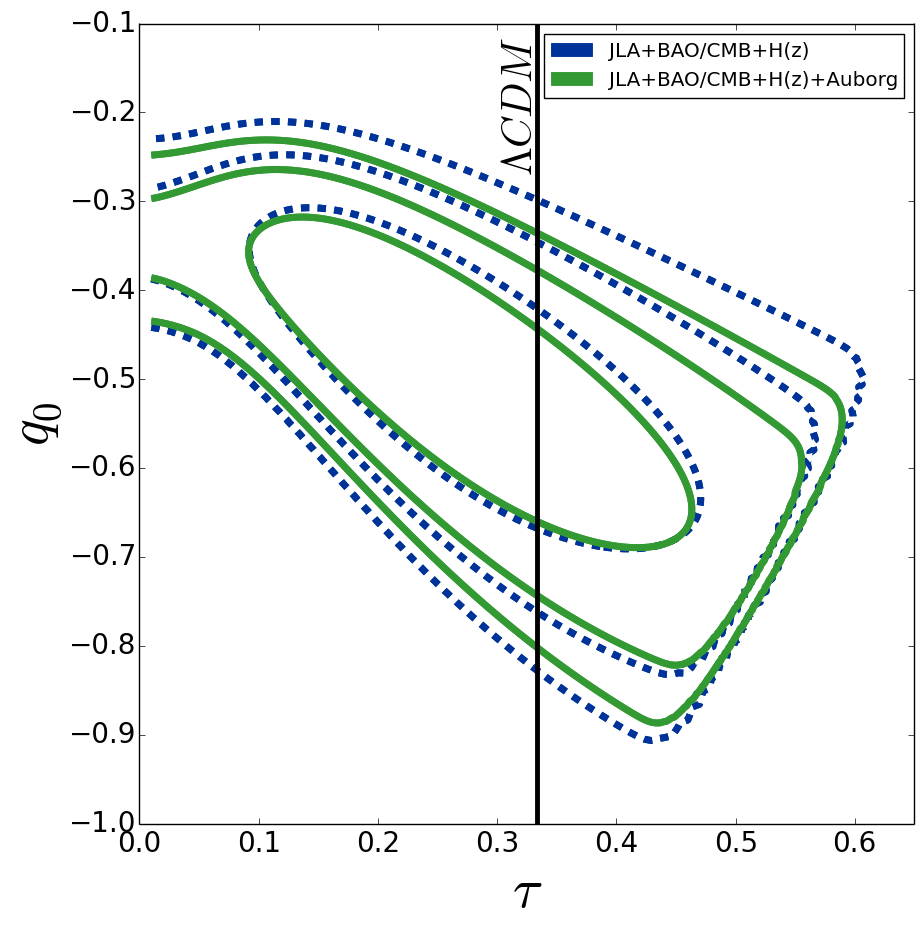}
\end{center}
\caption{\small{68\%, 95\% and 99\% confidence contours in the plane $q_0$ vs. $\tau$ 
for SN Ia + BAO/CMB(Planck) + H(z). The lines and color conventions are the same described in  Fig. \ref{fig4}}}
\label{fig6}
\end{figure*}

\begin{table}
\begin{center}
\begin{tabular}{| l | l | l | l | l | l | l |}
\hline
& \multicolumn{2}{|c|}{uniform prior} &\multicolumn{2}{|c|}{Aubourg \textit{et al.} prior} & \multicolumn{2}{|c|}{Riess \textit{et al.} prior} \\
\hline \hline 
& \multicolumn{1}{|c|}{$q_f=-1$} & \multicolumn{1}{|c|}{$q_f$ free} &\multicolumn{1}{|c|}{$q_f=-1$} & \multicolumn{1}{|c|}{$q_f$ free} &\multicolumn{1}{|c|}{$q_f=-1$} & \multicolumn{1}{|c|}{$q_f$ free} \\
\hline \hline
\multicolumn{1}{|c|} {$z_t$} & \multicolumn{1}{|c|} {$0.66_{-0.03(0.07)}^{+0.03(0.06)}$} & \multicolumn{1}{|c|}{$ 0.67_{-0.08(0.13)}^{+0.10(0.22)}$} & \multicolumn{1}{|c|} {$0.67_{-0.03(0.06)}^{+0.03(0.05)}$} & \multicolumn{1}{|c|}{$ 0.67_{-0.08(0.12)}^{+0.10(0.22)}$} & \multicolumn{1}{|c|} {$0.69_{-0.03(0.06)}^{+0.03(0.05)}$} & \multicolumn{1}{|c|}{$ 0.70_{-0.08(0.13)}^{+0.09(0.18)}$} \\
\hline
 \multicolumn{1}{|c|}{$\tau$} & \multicolumn{1}{|c|}{$0.33_{-0.04(0.07)}^{+0.04(0.09)}$} &\multicolumn{1}{|c|} {$0.26_{-0.10(0.18)}^{+0.14(0.26)}$} & \multicolumn{1}{|c|}{$0.33_{-0.03(0.06)}^{+0.04(0.08)}$} &\multicolumn{1}{|c|} {$0.26_{-0.10(0.17)}^{+0.14(0.26)}$} & \multicolumn{1}{|c|}{$0.31_{-0.03(0.06)}^{+0.03(0.07)}$}& \multicolumn{1}{|c|} {$0.24_{-0.09(0.16)}^{+0.13(0.25)}$} \\
\hline
 \multicolumn{1}{|c|}{$q_0$} & \multicolumn{1}{|c|}{-}&\multicolumn{1}{|c|}{ $-0.48_{-0.13(0.27)}^{+0.11(0.20)}$} & \multicolumn{1}{|c|}{-}&\multicolumn{1}{|c|}{ $-0.48_{-0.13(0.26)}^{+0.11(0.18)}$} & \multicolumn{1}{|c|}{-} & \multicolumn{1}{|c|}{ $-0.53_{-0.14(0.28)}^{+0.11(0.19)}$} \\
\hline
\end{tabular}
\caption{\small{Summary of the best-fit values for all parameters when using SN Ia + 
BAO/CMB(Planck) + H(z) ,
including the 68\% and 95\% confidence intervals. For each parameter we marginalized over all other ones with an uniform prior assuming their all possible values.}}
\end{center}
\label{tab1}
\end{table}

In Figs. \ref{fig4}, \ref{fig5} and \ref{fig6} we display 2D marginalized (an uniform prior in the 
third free model parameter) confidence contours ($68\%$, $95\%$ and $99\%$) in the 
$(z_t$, $\tau)$, $(z_t$, $q_0)$ and $(q_0$, $\tau)$ planes for SN Ia + BAO/CMB(Planck) + H(z) data. In all panels of the three figures, the blue dashed lines are the contours for the case in which marginalization over $H_0$ is performed with an uniform prior, while the red lines are the contours for the Riess \textit{et al.} prior and the green ones are for the Auborg \textit{et al.} prior. The 95\% C.L areas ratio relative to the Giostri \textit{et al.} results are $0.37$, $0.41$ and $0.44$, in the $(z_t$, $\tau)$ plane, $0.58$, $0.58$ and $0.71$ in the $(z_t$, $q_0)$ plane and $0.67$, $0.65$ and $0.75$, in the $(q_0$, $\tau)$ plane, for the Riess, Aubourg and uniform priors, respectively.

In Table 1 we show the best fit and the $68\%$ and $95\%$ confidence limits on each free parameter after marginalizing over all the other parameters with an uniform prior. In the cases we fix $q_f=-1$, there is no 
information in the Table on $q_0$ since, in these cases, only $z_t$ and $\tau$ are free parameters.
However, we can use Eq. (\ref{qfq0}) to obtain the expected values of $q_0$. We obtain, with 68\% of confidence level, $q_0=-0.54^{+0.05}_{-0.07}$ , $q_0=-0.56^{+0.06}_{-0.05}$  and 
$q_0=-0.59^{+0.06}_{-0.06}$ for the flat, Aubourg \textit{et al.} and Riess \textit{et al.}
priors, respectively.

\begin{figure*}[tbp]
\begin{center}
\includegraphics[scale=0.3]{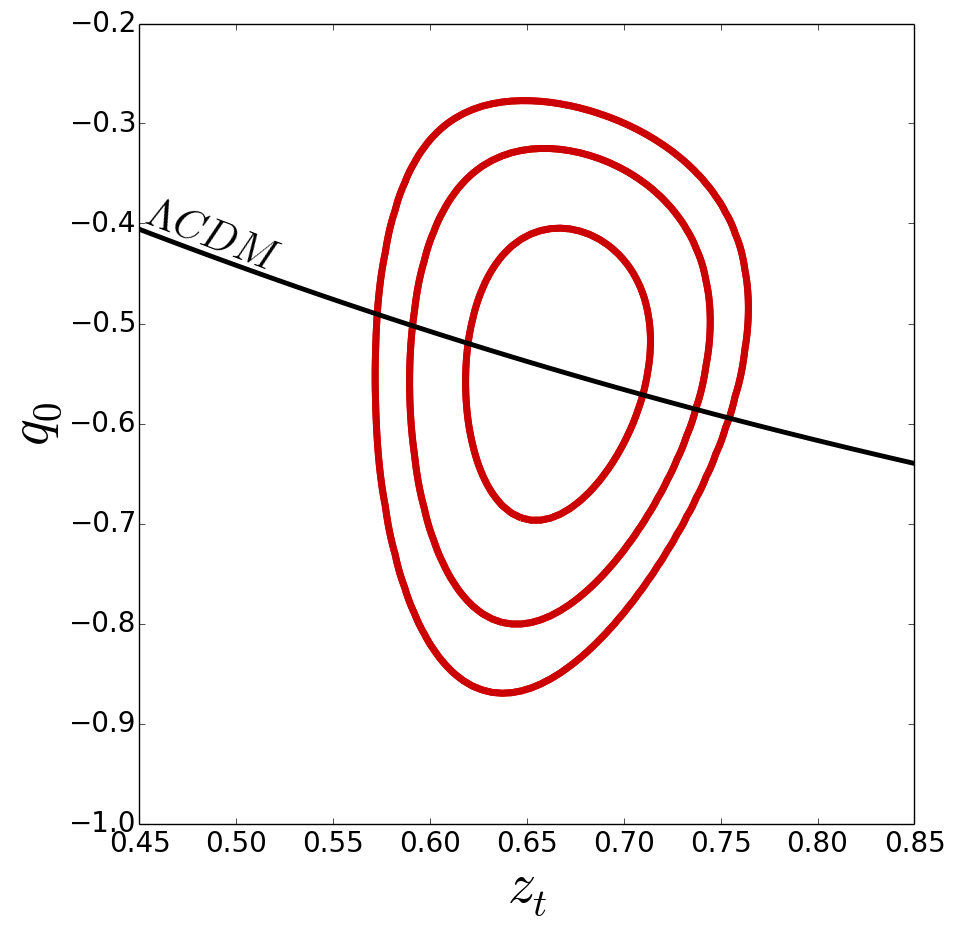}
\includegraphics[scale=0.3]{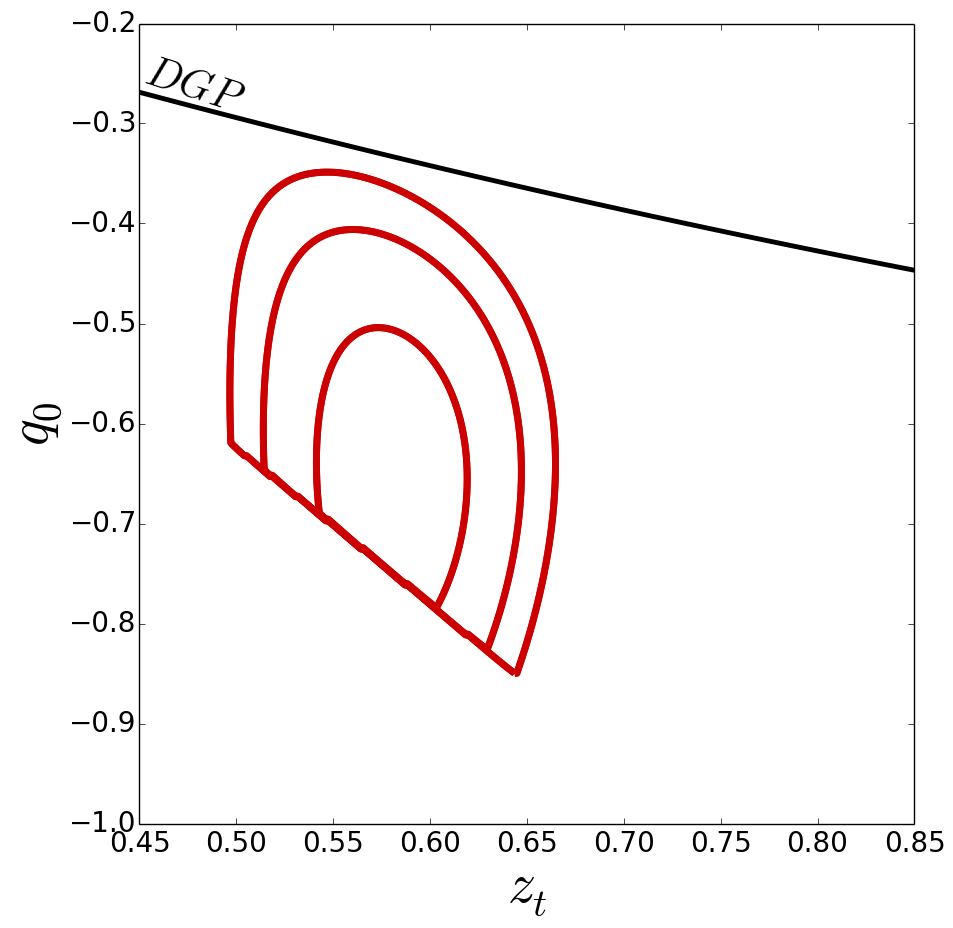}
\end{center}
\caption{\small{Sections, at $\tau=1/3$ (left panel) and $\tau=1/2$ (right panel) of the 68\%, 95\% and 99\% confidence 
surfaces in the parameter space 
($z_t$, $\tau$, $q_0$), marginalized over $H_0$ with an uniform prior. The black continuous 
line (left panel) 
corresponds to $\Lambda$CDM and the black dashed line (right panel) corresponds to 
the DGP model.}
}
\label{fig7}
\end{figure*}

\begin{figure*}[tbp]
\begin{center}
\includegraphics[scale=0.3]{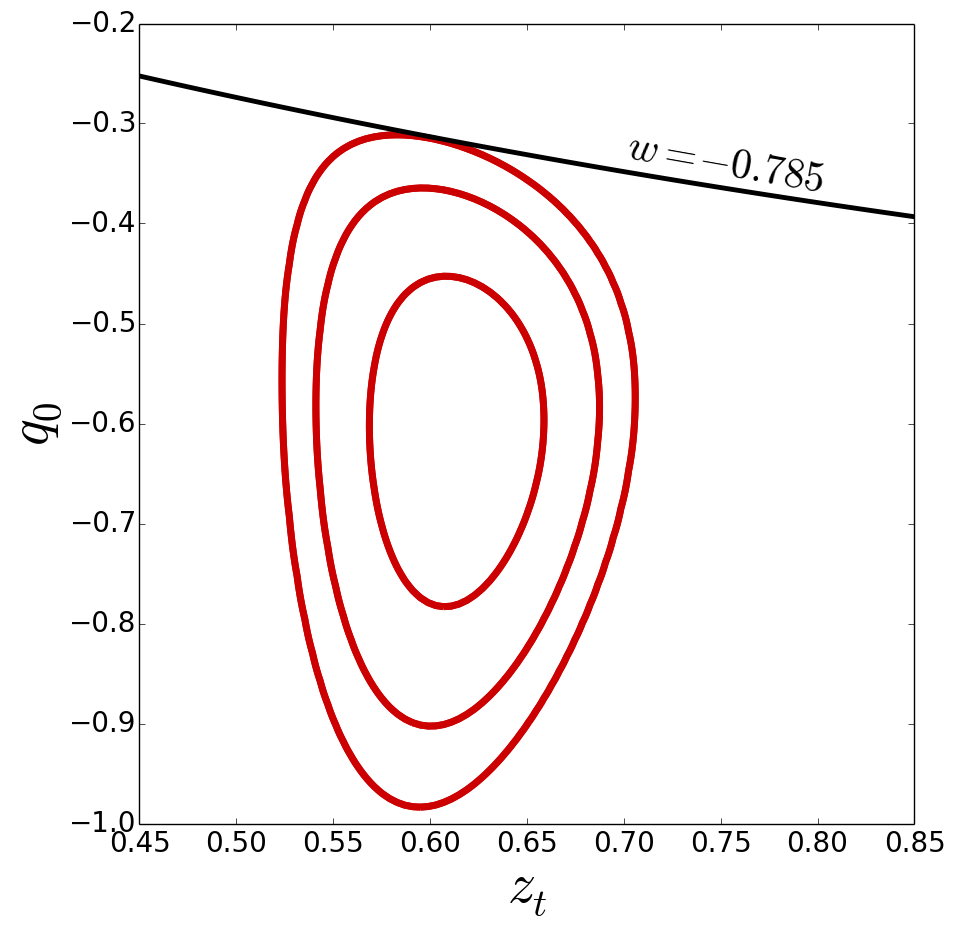}
\includegraphics[scale=0.3]{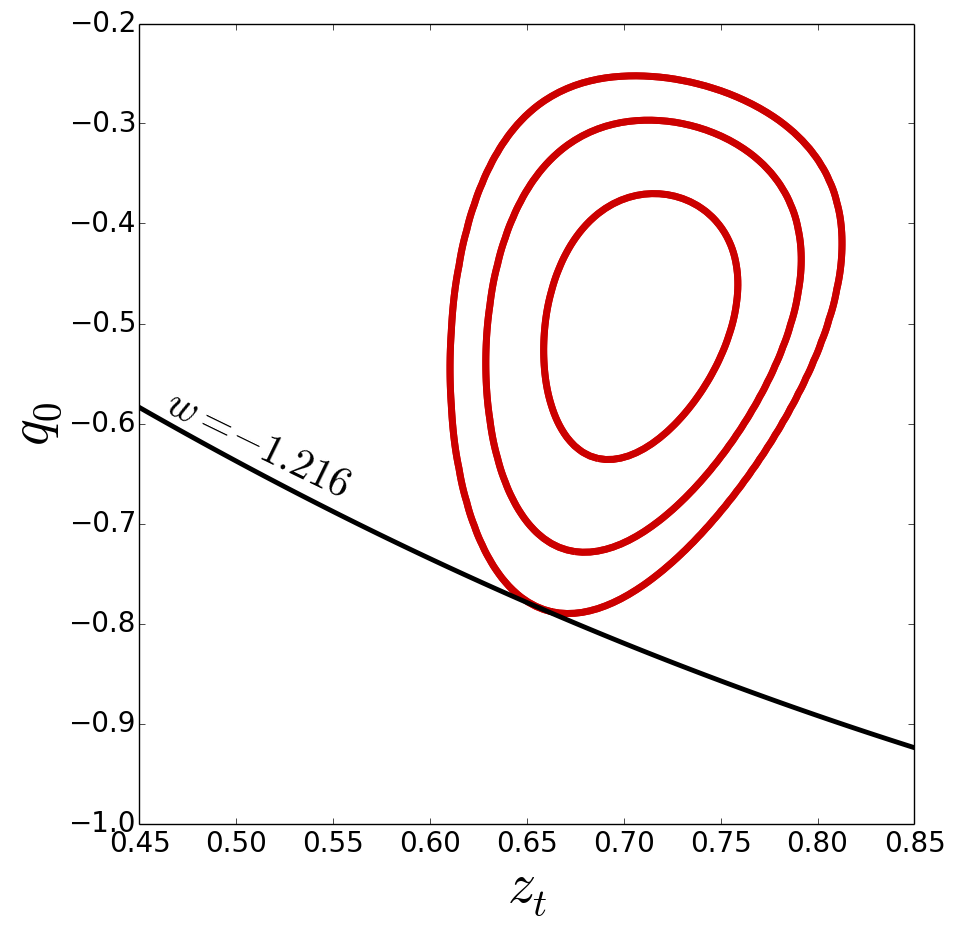}
\end{center}
\caption{\small{Sections $\tau=1/(3\times0.785)$ (left panel) and $\tau=1/(3\times1.216)$ (right panel) of the 68\%, 95\% and 99\% confidence 
surfaces in the parameter space 
($z_t$, $\tau$, $q_0$), marginalized over $H_0$ with an uniform prior. The black continuous line  in the left panel 
corresponds to $w$CDM with $w=-0.785$ while that one in the the right panel corresponds to $w=-1.216$.}
}
\label{fig8}
\end{figure*}

In the left panel of Fig.\,\ref{fig7} we show the $\tau = 1/3$ section of the 68\%, 95\% and 99\% confidence surfaces of the 
likelihood. We assume $q_f$ is free and an uniform prior marginalization of $H_0$. In the right panel, the $\tau = 1/2$ section is displayed. These two sections allow us to check directly the agreement between our results,
for three free parameters ($z_t$, $\tau$ and $q_0$),
and models with $\tau$ fixed, like $\Lambda$CDM ($\tau = 1/3$ and $q_f=-1$) and 
DGP ($\tau = 1/2$  and $q_f=-1$) 
models, since the marginalized plots in Figs. \ref{fig4}, \ref{fig5} and \ref{fig6} 
can be misleading.
We can see that $\Lambda$CDM, the black continuous line in the left panel, is in a very good agreement 
with the data, but the DGP model, the black dashed line in the right panel, is excluded at more than 99\% confidence level. It is worth mentioning that the cut in the contours in the right panel is due to the constraint on the parameter $\tau$, given by Eq. (\ref{tauconstr}), which ensures the transition from deceleration to acceleration, as discussed before.

In Fig.\,\ref{fig8} we show the $\tau=1/(3\times0.785)$ (left panel) and $\tau=1/(3\times1.216)$ (right panel) sections of the 68\%, 95\% and 99\% confidence surfaces of the likelihood. As in Fig.\,\ref{fig7} we assume that $q_f$ is free and an uniform prior marginalization over $H_0$. The black line in each panel corresponds to the flat $w$CDM model ($w=p/\rho=$ constant), with $w=-0.785$ (left panel) and $w=-1.216$ (right panel). Flat  $w$CDM are described by our parametrization if we assume $\tau=-\frac{1}{3w}$. We can see from Fig.\,\ref{fig8} that for $w$CDM our results indicate $-1.22 < w < -0.78$ at more than $99\%$ confidence level.

\section{Conclusions}
In this work we adopted a kinematical approach using a kink-like parametrization for the deceleration parameter to investigate the transition from cosmic deceleration to acceleration. 
To this end we considered recent SN Ia, BAO, CMB and H(z) observations.  As in \cite{ishida08} and \cite{giostri12}, we first considered the simple case in which $q_f=-1$ such that the models in this class have a final de Sitter attractor.  
With the new data the obtained constraints on the parameters are tighter but in good agreement with the previous ones. 

As compared with previous analysis \cite{giostri12} the first source for improvements is the SNeIa data used in this work. In  \cite{giostri12} it was considered 288 SNeIa while in the present work we consider 740. A naive point of view could expect a significant improvement in the constraints due to the higher statistics. However, in the present work correlations and systematics uncertainties were more properly considered. It is important to highlight the impact of the systematic errors in the supernova data \cite{betoule14}. It is well known that systematics tend to raise the confidence contours and it is in fact currently dominant for SNeIa. Therefore, the SNeIa treatment considered in this work is more robust. As we mentioned in Sec. \ref{sec:results}, we can get a 95\% C.L. contour areas (for $q_f=-1$) ratio, relative to the previous work, equal to 0.29 when using only BAO/CMB(Planck) data while we obtain 0.23 when we combine with SN Ia and H(z), which is an indication that the current SN Ia systematic errors are compensating most of the expected statistical enhancement. Regarding the BAO/CMB test we showed that, although we used more precise CMB information from Planck 2015 and WMAP9, the main source of improvements was the different BAO dataset used in this work. Here we disconsidered the older data at $z=0.2$ and $z=0.35$ and considered two more recent and precise measurements at $z=0.35$ and $z=0.57$. Of course, we expect to get tighter constraints once we have more and more precise BAO data in the future.

With the inclusion of H(z) data we were able to study the impact of the prior on $H_0$ in the results. We observed that assuming a Gaussian prior with the mean given by the measurement from Riess \textit{et al.} favors slightly lower values of $\tau$ and higher values of $z_t$, as compared to the Aubourg \textit{et al.} and the uniform prior marginalization. Although the $\Lambda$CDM model is in very good agreement with the results in all cases, the best fit in the Riess \textit{et al.} case is in slightly worse agreement. A possible explanation for this difference is that the Riess \textit{et al.} measurement is based in low redshift observations and seems to be less model dependent than the Aubourg \textit{et al.} one.  Although, in large part of our analysis we adopted the more conservative approach based on an uniform prior marginalization over $H_0$, we note that imposing Gaussian priors on $H_0$ is important to tighten the constraints on the parameters, mainly in the case in which $q_f=-1$.

In the case of three free parameters ($z_t$, $\tau$ and $q_f$) we also have improvements in the results as compared to previous work \cite{giostri12}. For instance, as can be seen in the right panel of Fig. \ref{fig7}, we are now able to exclude the DGP model at more than 99\% of confidence level even considering systematic errors in the SN analysis. As illustrated in the left panel of Fig. \ref{fig7}, we observe once more that flat $\Lambda$CDM model is in very good agreement with all the observables considered in this work. Finally, we also showed that for $w$CDM our results indicate $-1.22 < w < -0.78$ at more than $99\%$ confidence level.

\section*{\textbf{Acknowledgements}}

The authors would like to thank Ja\'{\i}lson Alcaniz and Rafael Tavares for helpful suggestions and the anonymous referee for constructive criticisms and comments that helped us to improve this work. M.V.S. thanks the Brazilian research agencies FAPERJ and CAPES for support. 

\appendix
\section{Marginalization over the nuisance parameters}
In this appendix we describe how the marginalizations, mentioned before, are performed. 

We start with the full $\chi^2$
\begin{equation}
	\chi^2 = \chi_{BAO/CMB}^2 + \chi_{SN}^2 + \chi_{H}^2,
\end{equation}
which corresponds to the likelihood
\begin{align}
	L &\propto \exp\left(-\frac{1}{2} \chi^2 \right) \\
	&= \exp\left(-\frac{1}{2} \chi_{BAO/CMB}^2 \right)\exp\left(-\frac{1}{2} \chi_{SN}^2 \right)\exp\left(-\frac{1}{2} \chi_{OHD}^2 \right),
\end{align}
that can be written as 
\begin{equation}
	L(H_0,M,\theta) = L_{BAO/CMB}(\theta)\; L_{SN}(H_0,M,\theta) \; L_{H}(H_0,\theta),
\end{equation}
where $\theta$ is a vector of the model parameters. In our case $\theta = \{z_t,\tau,q_i,q_f\}$. 

We now analytically marginalize over $H_0$ and $M$ such that
\begin{equation}
	L_{marg}(\theta) = \int L(H_0,M,\theta)\, d H_0 \,dM,
\end{equation}
where the integral is considered over all the possible values of the parameters.  
The first difficulty we find, when integrating the expression above, is that $L_{SN}$ and $L_{H}$ have a distinct functional dependence on $H_0$. The $\chi_{SN}^2$ is proportional to $(\log H_0)^2$ while $\chi_{H}^2$ is proportional to $H_0^2$. Fortunately this problem can be solved by making the following change of variable: $\{ H_0 , M \} \rightarrow \{ H_0 , \mathcal{M} = M + 5 \log H_0 \}$. So, the integral becomes
\begin{align}
	L_{marg}(\theta) &= \displaystyle\int L(H_0,\mathcal{M},\theta) \left| \frac{\partial(H_0,M)}{\partial(H_0,\mathcal{M})} \right| \,dH_0 \,d \mathcal{M}, \\
	&= \displaystyle\int L_{SN}(\mathcal{M},\theta) \;L_{H}(H_0,\theta)\; L_{BAO/CMB}(\theta) \,dH_0 \,d \mathcal{M}, \\
	&= L_{BAO/CMB}(\theta) \int L_{SN}(\mathcal{M},\theta) d\mathcal{M} \int L_{H}(H_0,\theta) dH_0, \\
	&= L_{BAO/CMB}(\theta) \;L_{SN,marg}(\theta) \;L_{H,marg}(\theta).
\end{align}
Therefore,
\begin{equation}
	\chi_{marg}^2 = -2 \log L_{marg} = \chi_{BAO/CMB}^2 + \chi_{SN,marg}^2 + \chi_{OHD,marg}^2 \;.
\end{equation}

We now perform each marginalization as follows
\begin{align}
	L_{SN,marg}(\theta) &\propto \int \exp\left( -\frac{1}{2} \chi_{SN}^2 \right) d\mathcal{M}
	&\propto \int \exp\left[ -\frac{1}{2} \left( A - 2\mathcal{M} B + \mathcal{M}^2 C \right) \right] d\mathcal{M}
	\label{snmarg}
\end{align}
where
\begin{align}
	A &= (\mu_b - 5 \log D_L (z_b,\theta))^t C_b^{-1} (\mu_b - 5 \log D_L (z_b,\theta)), \\
	B &= (\mu_b - 5 \log D_L (z_b,\theta))^t C_b^{-1} \mathbf{1}, \\
	C &= \mathbf{1}^t C_b^{-1} \mathbf{1}.
\end{align}
Solving the integral we get
\begin{equation}
	\chi_{SN,marg}^2 = A - \frac{B^2}{C} .
\end{equation}

Similarly we obtain for $L_{H,marg}$
\begin{align}
	L_{H,marg} &\propto \int \exp\left( -\frac{1}{2} \chi_{H}^2 \right) \,dH_0 \\
	&\propto \int \exp\left[ -\frac{1}{2} \left( A H_0^2 - 2 H_0 B \right) \right] \,dH_0 ,
\end{align}
where
\begin{align}
	A &= \sum_{i=1}^{N} \frac{ E^2 (z_i,\theta)}{\sigma_{H,i}^2 }, \\
	B &= \sum_{i=1}^{N} \frac{E (z_i,\theta) H_{obs,i}}{\sigma_{H,i}^2 },
\end{align}
and $E(z,\theta) = H_{th}(z,\theta)/H_0$. Thus,
\begin{equation}
	\chi_{H,marg}^2 = \log A - \frac{B^2}{A}.
	\label{Hmarg}
\end{equation}

In the case we consider a Gaussian prior over $H_0$, we have to work with the posterior
\begin{equation}
	P(H_0,\mathcal{M},\theta) = \Pi(H_0) L(H_0,\mathcal{M},\theta) \;,
\end{equation}
where $\Pi$ is the prior. Then the marginalized posterior is
\begin{align}
	P_{marg}(\theta) &= \int \Pi(H_0) L(H_0,\mathcal{M},\theta) dH_0 d\mathcal{M} \\
		&= L_{BAO/CMB}(\theta) \int L_{SN}(\mathcal{M},\theta) d\mathcal{M} \int \Pi(H_0) L_{H}(H_0,\theta) dH_0 \\
	&= L_{BAO/CMB}(\theta) L_{SN,marg}(\theta) P_{H,marg}(\theta).
\end{align}
where
\begin{align}
	P_{H,marg}(\theta) &= \int P_{H}(H_0,\theta) dH_0 \\
	&= \int \Pi(H_0) L_{H}(H_0,\theta) dH_0
\end{align}
and
\begin{align*}
	-2 \log P_{H}(H_0,\theta) &= \chi_{H}^2 - 2 \log \Pi(H_0) \\
	&= \sum_{i=1}^N \left[\frac{H_{th}(z_i,H_0,\theta) - H_{obs,i}}{\sigma_{H,i}}\right]^2 + \left( \frac{H_0 - H_{0,obs}}{\sigma_{H_0}} \right)^2 \\
	&= \sum_{i=1}^N \left[\frac{H_{th}(z_i,H_0,\theta) - H_{obs,i}}{\sigma_{H,i}}\right]^2 + \left[\frac{H_{th}(z=0,H_0,\theta) - H_{obs,0}}{\sigma_{H,0}}\right]^2 \\
	&= \sum_{i=0}^N \left[\frac{H_{th}(z_i,H_0,\theta) - H_{obs,i}}{\sigma_{H,i}}\right]^2.
\end{align*}
The expression above has the same form as the $\chi_{H}^2$ in (\ref{chi_hub}), but with one more data point, which corresponds to the prior. Therefore, we can proceed as before including only one more data point at $z=0$.

\end{document}